# TCP TRUNKING

H. T. Kung and S. Y. Wang
Division of Engineering and Applied Sciences
Harvard University
Cambridge, MA 02138, USA
email: kung@eecs.harvard.edu

July 1998

## ABSTRACT

A TCP trunk is an IP tunnel under TCP control, capable of carrying packets from any number of user flows. By exploiting properties of TCP, a TCP trunk provides elastic and reliable transmission over a network, and automatically shares the network fairly with other competing trunks. Moreover, by aggregating user flows into a single trunk flow, TCP trunking can significantly reduce the number of flows that the network needs to manage, thereby allowing use of simplified management to achieve improved performance. For example, when dealing with only a small number of TCP trunk flows, a router with a simple FIFO buffer can experience low packet loss rates.

A TCP trunk is a "soft" circuit in the sense that it requires no flow states to be maintained inside the network. Setting up a TCP trunk involves only configuring the two end nodes. This is in contrast with traditional methods of configuring circuits via signaling of network nodes.

A simple packet-dropping mechanism based on packet accounting at the transmitter of a TCP trunk assures that, when the trunk reduces its bandwidth in response to network congestion, user TCP flows carried by the trunk will reduce their bandwidths by the same proportion. Simulation results have demonstrated that TCP trunks can provide improved network performance to users, while achieving high network utilization.

## 1. INTRODUCTION

Methods of providing QoS for the Internet have been an active area of study for many years. A traditional approach is to use signaling to configure "circuits" of certain desired quality. Recent efforts in differentiated services [1] are aimed at providing QoS without signaling overheads.

In this paper we propose to use "TCP trunks" as a mean for assuring QoS. A TCP trunk over a network is an IP tunnel [2], which uses IP encapsulation to carry packets from any number of user flows. A TCP trunk differs from usual IP tunnels in that the transmission of data over the tunnel is controlled by TCP. Under TCP's congestion and flow control, the trunk is an elastic circuit in the sense that it will dynamically adjust its bandwidth to adapt to changing load conditions of the network. To set up a TCP trunk, only the two end nodes need to be configured and there is no need to configure intermediate nodes inside the network.

By exploiting TCP properties, multiple TCP trunks can fairly share a network. Through admission which will limit the number of TCP trunks sharing a network resource, and the number of user TCP flows sharing a trunk, such a trunk and user flow can be given a guaranteed minimum bandwidth. Under TCP control, the trunk and the user flow can expand automatically when extra bandwidth is available.

TCP trunking provides a solution to the problem [3] that packet drop rates of TCP connections sharing a bottleneck network link will increase as the number of these TCP flows increases. By aggregating multiple user flows into a single trunk flow via trunking, links and routers will only need to deal with a small number of trunk flows even when carrying a large number of user flows. This can also reduce the required buffer size, routing table size, and the number of route lookup operations, making the backbone routers scalable with a large number of user flows.

Moreover, TCP trunking provides a solution to the problem [4] that TCP connections with small windows are unfairly subject to TCP retransmission time-out during network congestion. (For example, interactive web sessions typically involve transfer of small files and thus use such TCP connections of small windows.) Through aggregation of user flows, a TCP trunk tends to transfer a larger data set and operate with a larger window than each individual user flow.

Configuring a TCP trunk requires only setting up TCP control at the two end nodes of the trunk. Nodes in the middle of the network need not be aware of the existence of the TCP trunk. (For minimum bandwidth guaranteeing mentioned above, the two end nodes of a TCP trunk may need to secure admission of the TCP trunk before the trunk connection is established. The admission can be obtained, for example, from a global controller responsible for limiting the total number of TCP trunks sharing network resources. No intermediate nodes inside the network need to participate in this admission process.)

In summary, taking advantage of knowledge and experiences accumulated over many years about TCP, TCP trunking provides a new type of layer-2 "circuits" for providing quality of service for layer-3 protocols such as IP. Because of its features described above, TCP trunking is distinctly different from traditional layer-2 methods such as ATM and Frame Relay, from tag-based approaches such as MPLS [5], and from generic IP tunnels [2].

To keep it short, this paper addresses only the basic ideas and rationale of TCP trunking. The paper will skip other considerations such as header encapsulation formats and various header compression methods. Future publications will address these other issues. The rest of the paper is organized as follows. Section 2 discusses some related work. Section 3 gives an overview of using TCP trunks in a network. Section 4 describes the buffer management at the transmitter of a TCP trunk. Section 5 presents simulation results for some typical usage scenarios of TCP trunking. Section 6 discusses reasons why TCP trunking can work well in a backbone network as demonstrated by the simulation results. Section 7 concludes this paper. Appendix A contains a list of FAQs to clarify some issues.

## 2. RELATED WORK

Conceptually, a TCP trunk is like a virtual path (VP) in an ATM network. Both of them aggregate multiple flows and implement some congestion control methods. However, they differ in many ways. TCP trunking approach uses a layer-4 TCP connection as a layer-2 link over IP networks. To set up a TCP trunk, only its two endpoints need to be configured. TCP trunking takes advantage of the widespread IP technology and thus can be easily and quickly deployed at any IP network. In contrast, an ATM VP is a pure layer-2 link using the ATM technology. To set up a VP, all switches on the VP's path need to be configured.

## 3. OVERVIEW OF TCP TRUNKING

This section gives a brief overview of TCP trunking. Figure 1(a) depicts an IP network with four router nodes. Figure 1(b) shows two TCP trunks, one from A to C and one from D to B. The users can use these trunks as if they were conventional leased lines or virtual circuits. Figure 1(c) exhibits two user TCP flows over tcp-trunk-1.

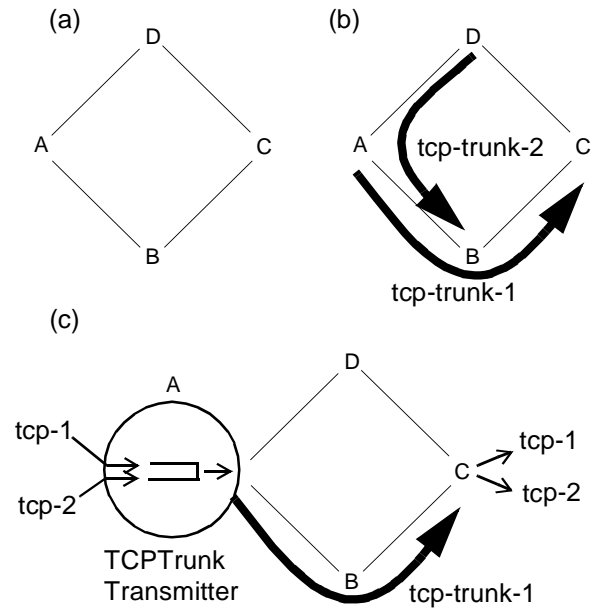

Figure 1: (a) IP network; (b) two TCP trunks over the physical links; and (c) two user flows (tcp-1 and tcp-2) over tcp-trunk-1.

A TCP trunk is a TCP connection. We call the source of this TCP connection the transmitter of the TCP trunk.

Consider, for example, tcp-trunk-1 of Figure 1(c). Packets arriving at the trunk transmitter from user flows tcp-1 ad tcp-2 are buffered in the socket buffer of the TCP connection for the trunk. Under the control of TCP, packets in the buffer are sent from node A to node C. When carried by the trunk, these packets are encapsulated with extra IP and TCP headers necessary for implementing the TCP connection for the trunk, and thus each packet will have two IP/TCP headers. Since a trunk is like a point-to-point link, usual header compression techniques [6] can be used to reduce an encapsulated packet's inside IP/TCP header (i.e., the packet's own IP/TCP header) overheads from 40 bytes to only 3~6 bytes. After header compression, a tiny header is prepended to the compressed packet before it is sent. This 2-byte tiny header specifies the length of the header-compressed packet so that the TCP trunk receiver will be able to detect packet boundary.

Multiple TCP trunks can share the same physical link. For example, tcp-trunk-1 and tcp-trunk-2 of Figure 1(b) share the same physical link from A to B. Each of these TCP trunks will adjust its share of the link bandwidth under TCP control.

When packets drop in a trunk due to congestion, the transmitter of the trunk will reduce its sending rate according to TCP's congestion control. When the buffer at the transmitter of the trunk becomes full, it will drop packets from the user flows over the trunk. When a packet from a TCP user flow is dropped, the user flow will in turn reduce its sending rate according to TCP's congestion control. To isolate UDP user flows, a separate buffer can be used in the TCP trunk transmitter.

## 4. BUFFER MANAGEMENT AT THE TRANSMITTER OF A TCP TRUNK

The buffer management at the transmitter of a TCP trunk addresses some of the challenging issues related to TCP trunking. These include interaction of the two levels (i.e., trunk and user levels) of TCP congestion control as well as fair and efficient use of the trunk by its user flows. We assume in this section that all user flows are TCP flows.

A TCP trunk relies on the TCP fast retransmit and recovery mechanism [7] to adjust its bandwidth in response to the congestion condition of the network. Roughly speaking, upon receiving three duplicated ACKs (as an indication of a probably lost packet due to network congestion), the transmitter of a TCP trunk will reduce its transmission rate by one half. As explained in the paragraph below, this will cause user flows on the trunk to drop their packets at the transmitter of the trunk. To achieve high trunk utilization and fairness among user flows on the trunk, it is critical that proper packet dropping policy is applied to these flows at the trunk transmitter.

Consider the situation when a TCP trunk drops a packet at some router on the trunk. (This lost "trunk" packet is an encapsulated "user" packet of some user flow.) After recognizing this packet loss, the trunk transmitter will reduce its sending rate. In the meantime, any user flow on the trunk may not necessarily experience any packet loss. It will thus continue transmitting at the same or increased rate, in spite of the fact that the underlying trunk has already reduced its rate. The queue of the user flow at the trunk transmitter will therefore build up, until some time after packets from the user flow are dropped at the queue due to queue overflow, and the packet dropping triggers the sender of the user flow to reduce its sending rate.

Ideally, when the trunk reduces its bandwidth by some factor, we would want all the active user flows over the trunk to also reduce their bandwidths by the same factor. We use the following three principles to provide an approximate solution for achieving this goal:

- P1. All user flows share a buffer of size about RTTup*TrunkBW, where RTTup is an upper estimate of RTTs of user flows, and TrunkBW is the target peak bandwidth for the TCP trunk. This buffer is to hide the control latency of user flows beyond that of the trunk, as explained above. More precisely, when the TCP trunk reduces its sending rate by one half using fast retransmit and recovery, the number of in-flight user packets is reduced from RTTup*TrunkBW to RTTup*TrunkBW/2. This implies that the number of user packets which may need to be queued at the TCP trunk transmitter is at most RTTup*TrunkBW - RTTup*TrunkBW/2 = RTTup*TrunkBW/2. Therefore, a buffer of size RTTup*TrunkBW is large enough to hold user packets without packet dropping when the TCP trunk reduces its sending rate by one half.

- P2. The buffer occupancy at the TCP trunk transmitter is constantly maintained at a value lower than a threshold. When a packet arrives at the buffer and the buffer occupancy is higher than the threshold, the arriving packet will be dropped with a probability proportional to its buffer occupancy at the TCP trunk transmitter. In particular, arriving packets will always be dropped when the buffer is full. This packet-dropping policy is similar to that of RED [8].

- P3. After dropping a packet from a user flow, the trunk transmitter will try not to drop another packet from the same user flow, until the user flow has recovered from this packet loss by fast retransmit and recovery. Note that for the user flow, dropping a packet will cause it to reduce its sending rate by one half. This rate reduction matches the rate reduction of the underlying trunk. Additional packet drops from the same user flow are likely to cause unnecessary TCP retransmission time-outs. Therefore, the goal here is to try to make TCP fast retransmit and recovery work every time when a user packet is dropped, for all user TCP flows.

We use a simple per-flow packet accounting method to implement the P3 principle above. The trunk transmitter will estimate the total number X of packets that can be sent by a user TCP flow source between the time it reduces its sending rate by one half and the time its sending rate is about to ramp up to its previous sending rate when its packet was dropped. We use this number X to set a threshold K, which will be the minimum number of packets from the TCP flow that should be forwarded without being dropped, before any packet from the same flow will get dropped again.

Suppose that the user TCP flow's congestion window has w packets when fast retransmit is triggered. The number X of packets sent during the fast retransmit and recovery period is roughly w/2+(w/2+1)+(w/2+2)+....+w = (3/8)*w^2. For example, when w is 10, X = 37. Since we do not want to drop another packet from this user flow too soon, the threshold K should not be too small. On the other hand, if K is set to be too large, the exemption period, when the flow can keep growing its sending rate beyond its fair share of the available bandwidth, could be too long. This would increase the steady-state buffer occupancy. For the simulation runs reported in this paper, the value of K is set to be X/2. The performance results are found to be not sensitive to the precise value of K.

The transmitter of the TCP trunk calculates values X and thus K as follows. The product of the TCP trunk's current bandwidth and an RTT estimate for the trunk is used to approximate the congestion window size W for the TCP truck flow. By tracking the number N of active user flows on the TCP trunk, the congestion window size of each active user flow is estimated to be W/N. Thus the threshold K is X/2 = (3/8)*(W/N)^2*(1/2). This value of K will be used for every active user flow when employing the buffer management scheme mentioned above.

## 5. SIMULATION OF TCP TRUNKS

We have used a TCP/IP network simulator [9] to study performance of TCP trunks. This simulator uses real-life BSD 4.4 networking code to send, forward, and receive TCP/IP traffic. Another feature of the simulator is that standard UNIX APIs are provided on every simulated node. This allows application programs to be developed and run on any node in a simulated network. We use this capability to develop the TCP trunk transmitter and receiver programs. The TCP trunk transmitter will intercept user flows' raw packets, and transmit them on a TCP trunk via a TCP socket. The TCP trunk receiver will receive these packets via a TCP socket and send these raw packets onto a link via a raw socket.

The following two subsections present simulation results for two usage scenarios of TCP trunking.

### 5.1. Simulation Scenario I: Protection for Interactive Web Users

This suite of simulation results show that TCP trunking can provide protection for interactive web users in the sense that these users can receive their fair share of the available bandwidth and therefore avoid unnecessary time-outs.

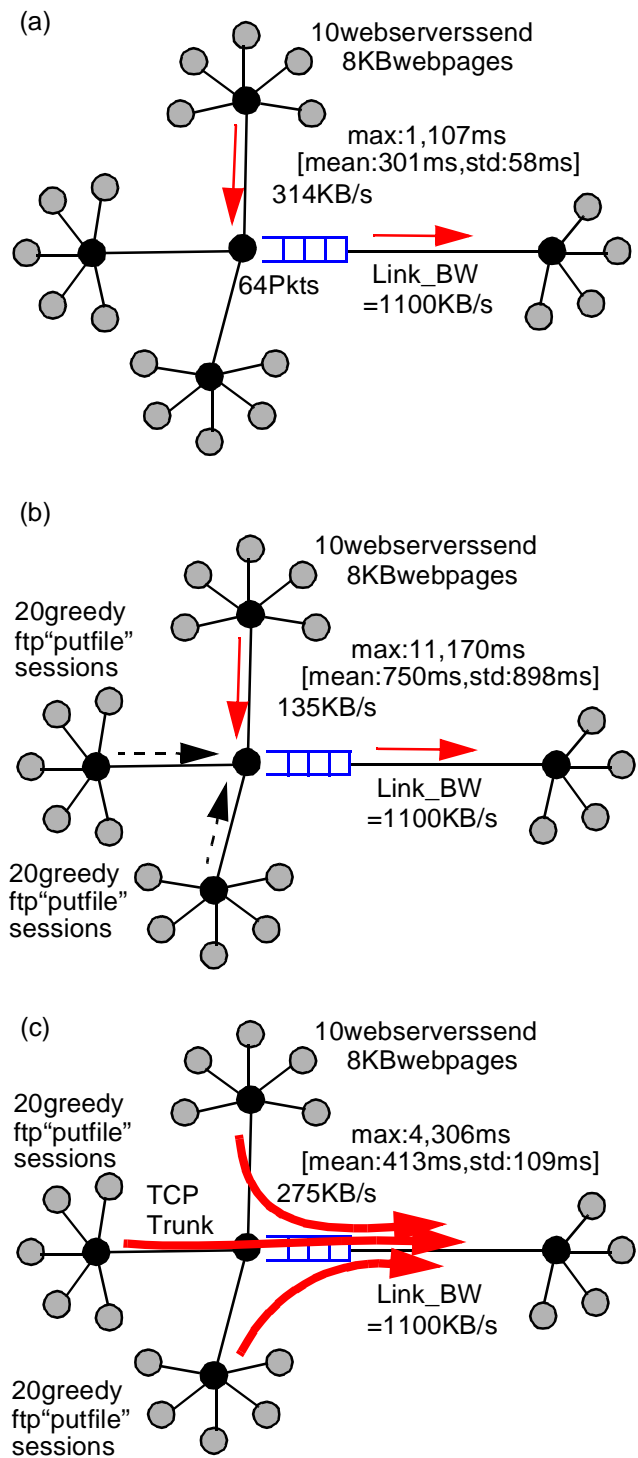

Figure 2: Website throughput and transfer (a) under no competing ftp traffic; and (b) under competing ftp traffic. (c) Web side performance for load (b) when three TCP trunks, one for each site, are used.

Consider the configuration depicted in Figure 2(b). On the middle router where traffic merges, there are many short-lived web transfers coming from an input port (a site) to compete for an output port's bandwidth (1100 KB/sec) with other long-lived greedy ftp transfers that come from two other input ports (sites).

Figure 2(a) shows that when there are only short-lived, 8KB web transfers in the network, the offered load uses only 314 KB/sec bandwidth and is less than the site's fair share (1100/3 KB/sec) of the output port's bandwidth. (The offered load is limited, because TCP windows for these web transfers never ramp up significantly, due to the small 8KB size of the transfers.) Also, the request-response delays for these short-lived web transfers are small and predictable. The mean delay, maximum delay, and the standard deviation of the delays are 301 ms, 1,107 ms, and 58 ms, respectively.

Figure 2(b) shows that after long-lived greedy ftp transfers ("putfile" sessions) are introduced into the network, the short-lived web transfers can only achieve 135 KB/sec bandwidth in aggregate, which is much smaller than its fare share (1100/3 KB/sec). The mean delay, maximum delay, and the standard deviation of the delays increase greatly and become 750 ms, 11,170 ms, and 898 ms, respectively. This means that the short-lived web transfers are very fragile and encounter more time-outs than before. As a result, they cannot receive their fair share of the bandwidth of the bottleneck link when competing with long-lived greedy ftp transfers.

Figure 2(c) shows that when a TCP trunk is used for each site to carry the site's aggregate traffic, the bandwidth used by the short-lived web transfers increases to 275 KB/sec. The mean delay, maximum delay, and the standard deviation of the delays also improve greatly and become 413 ms, 4,306 ms, and 109 ms, respectively. These performances are close to the best performances that the short-lived web transfers can achieve when a fair allocation of bandwidth is allocated to them.

## 5.2. Simulation Scenario II: Local Control to Assure QoS

TCP trunking allows a site to control its offered load into a backbone network so that the site can assure some QoS for its packets over the backbone. This is in contrast to the current Internet situation where all sites experience the same QoS regardless of their offered loads.

Consider the configuration depicted in Figure 3(a). In the first part of the simulation, aggregate traffic from eight user sites is merged on a router connected to a cluster of servers.

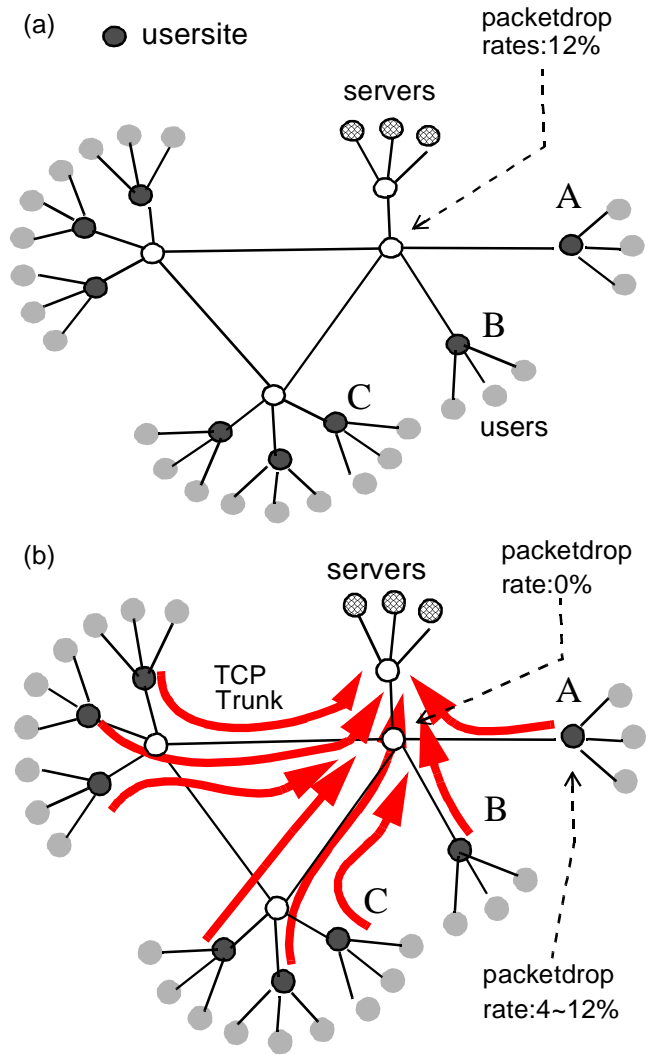

Figure 3: Users from eight site transfer files to a server site, (a) without TCP trunks, and (b) with TCP trunks one for each user site.

The aggregate traffic consists of 280 greedy ftp "putfile" transfers from the users to the servers. Since the bottleneck link bandwidth is set to be only 10 Mbps and the buffer on that link is relatively small (130 packets) with respect to the 280 TCP connections, a very high packet drop rate of 12% is observed on that link.

In contrast, suppose now that a TCP trunk is employed to carry each site's aggregate traffic, as depicted in Figure 3(b). Since there are only 8 competing TCP trunk connections on the bottleneck link, given the same size of the shared buffer, packet drop rates on the bottleneck link decrease significantly (in fact, they are almost 0%).

Packet dropping rates, however, remain high on the TCP trunk transmitters. In fact, the dropping rates (between 4% and 12%) can be as high as before. (Each TCP trunk needs to carry only one-eighth of the total user TCP flows, but its fair share of the available bandwidth of the bottleneck link is also only one-eighth). Thus, from the viewpoint of each user TCP flow, end-to-end packet drop rates before and after using TCP trunking may be about the same. However, from the whole system's point of view, TCP trunking has helped restrict packet drops to the edge nodes which host transmitters of TCP trunks, so that the backbone network need not waste its bandwidth for forwarding packets which will be dropped later.

Perhaps more importantly, to decrease end-to-end packet drop rates for its user flows, a local site can increase the size of the buffer in the transmitter of its TCP trunk. Also, to improve the fairness among its user flows in their uses of the trunk, a local site can choose to use sophisticated buffer management schemes, e.g., [3, 8]. These configurations can all be decided and carried out locally within each site, without concerns about the other sites and the backbone network.

In the second part of the simulation, we focus on traffic from sites A, B and C on the right of the Figure 3(b). The offered load of site A to the backbone network is 30 greedy ftp file transfers, that of site B's is 3 greedy ftp file transfers, and that of site C is 30 small ftp sessions modeling web traffic. For each of these small ftp sessions from site C, the sender will continuously transfer a new short file of 8KB when the previous transfer is completed. Again, these 3 sites' offered load plus the other sites' offered load make up 280 ftp transfers. These 280 connections will compete for the bandwidth of the bottleneck link connected to the server site.

In addition, we added a short transfer session, like any of the 30 small ftp transfer sessions from site C, to each site to probe the end-to-end request-response delay experienced by the site. The simulation results show that, when TCP trunk is not used, regardless of a site's offered load, every site experiences the same mean request-response delay (about 3.78 sec). When TCP trunk is used, the observed mean request-response delay for site A, B, and C is 3.83 sec, 0.82 sec, and 0.71 sec, respectively.

These results suggest that via TCP trunking, QoS over a backbone can be assured locally by local admission control at a site without having to cooperate with other sites. For example, from the simulation results, we see that when the offered load from a site (site B or C) is no more than its fair share of the available bandwidth, the resulting request-response delay is good and predictable. Therefore, with TCP trunking, each site will have the incentive to use only its fair share of the available bandwidth, and to restrain itself from dumping excessive traffic into the backbone network.

## 6. FAIR BANDWIDTH ALLOCATION AMONG TCP TRUNKS

When a TCP trunk carries a site's aggregate traffic over a backbone network, its achieved bandwidth determines the site's allowed outgoing bandwidth to the backbone. Therefore, a fair bandwidth sharing among competing TCP trunks over the backbone is important.

Per-trunk queueing and scheduling can be used in the backbone network to ensure that each TCP trunk receives its fair share of the network bandwidth. This is relatively easy to achieve, since there can be only a small number of TCP trunks.

Sometimes multiple TCP flows can still share a network link fairly and efficiently even when only FIFO queueing and scheduling is used. This will be the case when, for these flows, the RTTs do not differ significantly, and fast retransmit and recovery work most of the time. To allow for fast retransmit and recovery, each TCP flow must have a chance to grow up its congestion window to a sufficiently large size. This requirement can be met when the following two conditions hold:

- C1: Each TCP flow is a relatively long transfer, involving, for example, one hundred or more packets, so that its TCP congestion window can ramp up to at least five packets. It is well-known that the TCP congestion window needs to be at least this size before TCP fast retransmit and recovery can work.

- C2: The shared network bandwidth and buffer is large enough to allow each TCP's congestion window to grow up to at least five packets. This will allow the TCP fast retransmit and recovery mechanism to work as noted in C1 above. (The buffer size can be reduced under flow-aware schemes such as FRED [3], and modified TCP sender algorithm [4].)

For TCP trunk flows, C2 can be assured via TCP trunks admission process. By limiting the number of TCP trunks sharing a buffer and a link, each TCP trunk can grow its congestion window to a guaranteed number of packets. Satisfying C1 can be expected because a TCP trunk normally aggregates traffic from many user flows. This suggests that a simple buffer management scheme such as FIFO can work well, when the competing TCP trunks do not have significant disparities in RTT. This reasoning has been confirmed by the simulation results presented in Section 5.

## 7. CONCLUDING REMARKS

TCP trunking can meet various network performance goals demanded by applications. For example, TCP trunking can be used in packet-based transport networks [10] and in virtual private networks [11] to provide layer-2 services. TCP trunking's tunneling capability is similar to that of "L2TP" [12], which is commonly used in virtual private network products, but provides more sophisticated flow control than "L2TP" in dealing with network congestion.

## Appendix A. Frequently Asked Questions (November 1998)

Q1. Has TCP trunking been implemented and tested?

A1. Yes. A user-level implementation has been running on FreeBSD 2.2.7 since July 1998. With a 300-MHz Pentium PC, we observe that packets carried on a TCP trunk incur about 0.2 ms additional latency. Performance measurements are being carried out on a testbed network at Harvard.

Q2. Is TCP trunking approach scalable? It seems that there can be many TCP trunks that need to be configured in the network?

A2. TCP trunking is designed as an alternative to fixed-bandwidth layer-2 circuits for transport networks. Usual approaches of reducing #trunks to be managed, such as traffic aggregation and hierarchy, are also applicable to TCP trunking. We are working on these issues.

Q3. How does TCP trunking handle TCP and UDP traffic?

A3. Packets will be classified before they enter a TCP trunk. UDP and TCP packets will be carried on separate TCP trunks. By doing this, user TCP flows can get their fair bandwidth shares in the presence of aggressive user UDP flows. In general, packets will be classified into different service classes each to be carried by a dedicated TCP trunk.

Q4. If my site first uses a TCP trunk to carry my aggregate traffic but other sites don't, my site's will suffer. This is because a site using more TCP connections to the backbone will generally achieve more bandwidth than a site using just one TCP connection. Can there be incentive for a site to be an early TCP trunking adopter?

A4. ISPs will need to be sensitive on this issue. Being friendly to the network, customers using TCP trunks may be offered with some price discounts. Customers not using TCP trunks may have their traffic all aggregated in a few public TCP trunks.

Q5. Is the extra TCP/IP header overhead required in TCP trunking significant?

A5. Given that the normal packet size is 1500 bytes and that TCP/IP header needs 40 bytes, the overhead is only about 2.6%. Furthermore, the IP/TCP header inside an encapsulated packet can be compressed from 40 bytes to 3~6 bytes using typical compression techniques as noted earlier in the paper.